\documentclass[pra,a4paper,superscriptaddress,twocolumn,amsmath,amssymb]{revtex4-1}
\pdfoutput=1
\usepackage{graphicx}
\graphicspath{{./Figs/}}
\usepackage{color}
\usepackage{bm}
\usepackage{dsfont}
\usepackage{epstopdf}

\usepackage{verbatim}
\newcommand {\avg}[1]{\langle #1 \rangle}

\newcommand{\ket}[1]{\left| #1 \right\rangle}

\newcommand{\tone}{$T_1$}

\begin{document}

\title{Charge dynamics and spin blockade in a hybrid double quantum dot in silicon}
\author{Matias Urdampilleta	}
\email{m.urdampilleta@ucl.ac.uk}
\affiliation{London Centre for Nanotechnology, University College London, London WC1H 0AH, UK}

\author{Anasua Chatterjee }
\affiliation{London Centre for Nanotechnology, University College London, London WC1H 0AH, UK}

\author{Cheuk~Chi Lo }
\affiliation{London Centre for Nanotechnology, University College London, London WC1H 0AH, UK}
\affiliation{Dept.\ of Electronic \& Electrical Engineering, University College London, London WC1E 7JE, UK}

\author{Takashi Kobayashi }
\affiliation{Centre for Quantum Computation and Communication Technology,
School of Physics, University of New South Wales, Sydney NSW 2052, Australia}

\author{John Mansir }
\affiliation{London Centre for Nanotechnology, University College London, London WC1H 0AH, UK}

\author{Sylvain Barraud}
\affiliation{CEA, LETI, Minatec Campus, F-38054 Grenoble, France}

\author{Andreas~C. Betz}
\affiliation{Hitachi Cambridge Laboratory, J. J. Thomson Avenue, Cambridge CB3 0HE, U.K}

\author{Sven Rogge }
\affiliation{Centre for Quantum Computation and Communication Technology,
School of Physics, University of New South Wales, Sydney NSW 2052, Australia}

\author{M.~Fernando Gonzalez-Zalba}
\email{mg507@cam.ac.uk}
\affiliation{Hitachi Cambridge Laboratory, J. J. Thomson Avenue, Cambridge CB3 0HE, U.K}

\author{John~J.~L. Morton}
\affiliation{London Centre for Nanotechnology, University College London, London WC1H 0AH, UK} 
\affiliation{Dept.\ of Electronic \& Electrical Engineering, University College London, London WC1E 7JE, UK} 


\begin{abstract}
Electron spin qubits in silicon, whether in quantum dots or in donor atoms, have long been considered attractive qubits for the implementation of a quantum computer due to silicon's  ``semiconductor vacuum" character and its compatibility with the microelectronics industry. While donor electron spins in silicon provide extremely long coherence times and access to the nuclear spin via the hyperfine interaction, quantum dots have the complementary advantages of fast electrical operations, tunability and scalability. Here we present an approach to a novel hybrid double quantum dot by coupling a donor to a lithographically patterned artificial atom. Using gate-based rf reflectometry, we probe the charge stability of this double quantum dot system and the variation of quantum capacitance at the interdot charge transition. Using microwave spectroscopy, we find a tunnel coupling of 2.7~GHz and characterize the charge dynamics, which reveals a charge $T_2^*$ of $200\,$ps and a relaxation time $T_1$ of $100\,$ns. Additionally, we demonstrate spin blockade at the inderdot transition, opening up the possibility to operate this coupled system as a singlet-triplet qubit or to transfer a coherent spin state between the quantum dot and the donor electron and nucleus.
\end{abstract}

\maketitle
\section{Introduction} 
 Interest in donor-based spin-qubits in silicon has been motivated by their exceptionally long electron spin coherence times, exceeding one second in isotopically enriched  $^{28}$Si \cite{Tyryshkin2012}. Additionally the donor electron spin can be a gateway to access the donor nuclear spin, which has longer coherence times \cite{Steger2012}, even at room temperature \cite{Saeedi2013} and the potential to serve as a quantum memory \cite{Morton2008}. Moreover, the single shot read-out of single electron \cite{Pla2012} and nuclear \cite{Pla2013} spins, a milestone for donor-based quantum computing, has been recently demonstrated in nanoelectronic silicon devices.

On the other hand, artificial atoms such as electrostatically defined quantum dots offer complementary advantages as qubits, notably in their tunability~\cite{Veldhorst2014}, flexible coupling geometries~\cite{Shulman2012}, and opportunities for fast electrical control of spin~\cite{Foletti2010}. It is therefore attractive to investigate the possibility of \emph{hybrid} architectures which brings together the advantages of these two systems by coupling a quantum dot to a donor atom~\cite{Schenkel2011,FernandoGonzalez-Zalba2012}. Such a double dot could take advantage of fast spin manipulations using gate voltage to form a hybrid singlet-triplet qubit coupled to the long-lived quantum memory offered by its nuclear spin. In addition, this hybrid architecture could be used to create spin buses with quantum dot chains to mediate quantum information stored in donor qubits over long distances~\cite{Friesen2007}. 

In this work, we present measurements taken on a silicon nanowire transistor indicating a coupled system formed of a single phosphorus atom and a quantum dot. We show that this hybrid system behaves as a double quantum dot, and characterise the high-frequency admittance of the coupled system by RF gate-based sensing. Quantum capacitance changes at the interdot charge transition (ICT) and microwave (MW) spectroscopy allow us to characterize the tunnel coupling and the charge dynamics. Finally, we demonstrate spin blockade effects between singlet and triplet states by applying a magnetic field which changes the resonator response at the ICT. These results demonstrate the potential of a donor-dot system as a new singlet-triplet qubit.

\section{Characterization of the device} 
\subsection{Device fabrication and measurement setup}
\begin{figure}[t]%
\includegraphics[width=\columnwidth]{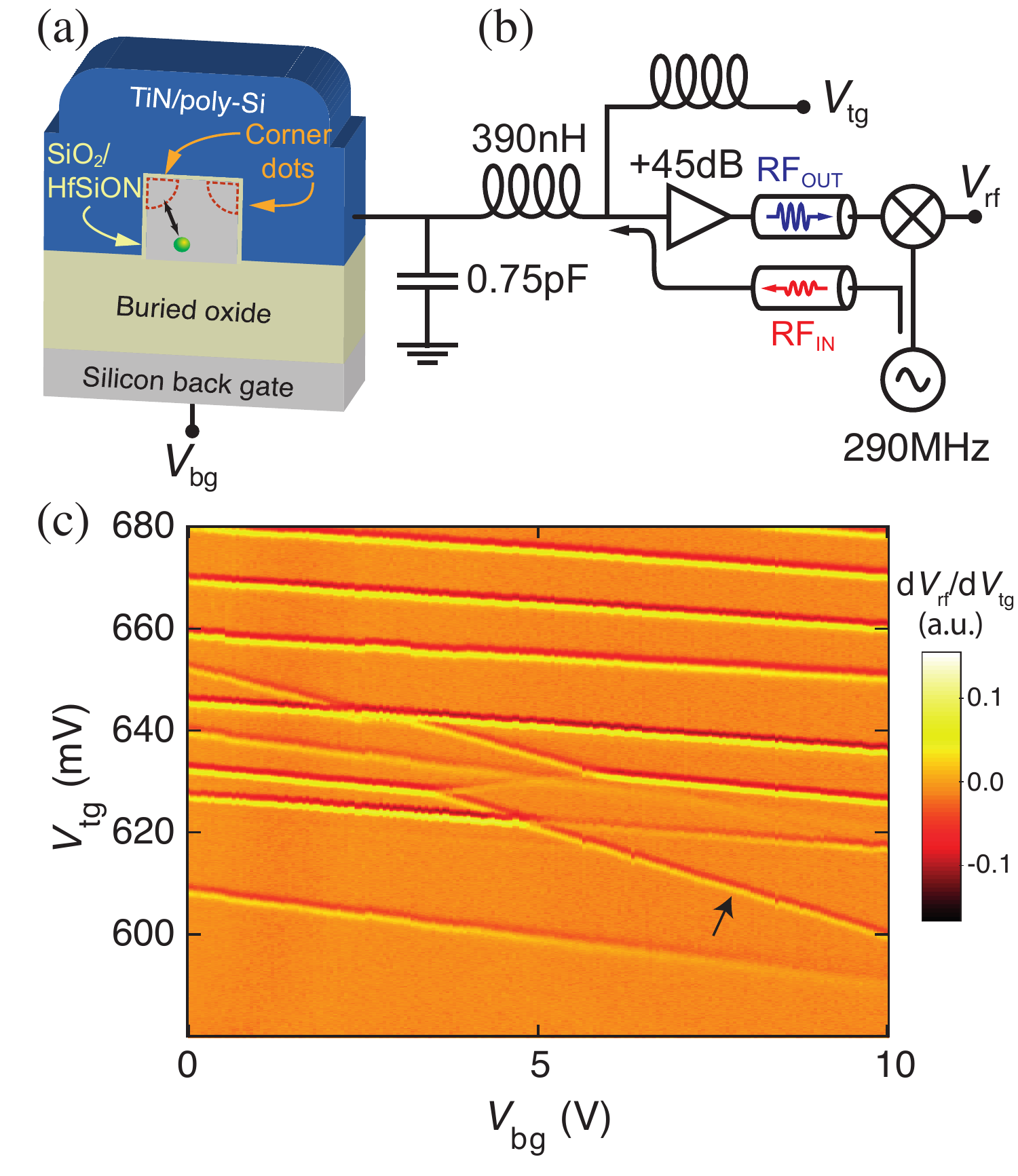}%
\caption{(a) A sketch of the device cross-section shows quantum dots localized in the top corners of the nanowire and a dopant located deeper in the channel. (b) In our rf reflectometry measurements, an rf signal is sent through a directional coupler to a resonator made from the device capacitance, parasitic capacitance and surface-mounted inductor. The reflected signal is amplified at low temperature, and demodulated using a reference signal to give $V_{\rm rf}$. (c) The derivative of the demodulated resonator response as a function of $V_{tg}$ and $V_{bg}$ gives the charge stability diagram. Transitions with weak $V_{bg}$ dependence are attributed to changes in the charge occupancies of the two distinct (and uncoupled) corner dots. The charge transition indicated with a black arrow is attributed to a phosphorus atom in the bulk of the nanowire.}
\label{fig:Sample_stability}%
\end{figure}
The nanowire transistor device, sketched in Fig.~\ref{fig:Sample_stability}(a), is completely CMOS industry compatible and is fabricated from an SOI substrate with a 145\,nm buried oxide and a silicon layer doped with phosporus at a concentration of 5$\times$10$^{17}\,$cm$^{-3}$. The doped  silicon layer is etched to create a $200\,$nm long and $30\,$ nm wide nanowire by means of deep-UV lithography. A $30\,$nm wide wrap-around top gate is defined using a SiO$_2$($0.8\,$nm)/HfSiON($1.9\,$nm) stack for the gate dielectric followed by TiN($5\,$nm)/poly-Si($50\,$nm) as the top gate material. The source and drain are self aligned and formed by ion implantation after the  deposition of $20\,$nm long Si$_3$N$_4$ spacers. 

Measurements are performed at the base temperature of an Oxford Instrument Triton 200 cryogen-free dilution refrigerator ($30\,$mK). High sensitivity charge detection is achieved by radiofrequency reflectometry on a tank circuit composed of a surface mounted inductance (390\,nH), a parasitic capacitance to ground ($0.75\,$pF) and the device capacitance between the transistor top gate and the channel \cite{Gonzalez-Zalba2015}. Rf-reflectometry is performed close to resonance frequency ($294\,$MHz), with the input power set to $-85\,$dBm and the reflected signal amplified by a low noise cryogenic amplifier anchored at $4\,$K. The signal is further amplified and demodulated at room temperature as shown in Fig.~\ref{fig:Sample_stability}(b). An on-board bias tee is used to apply both DC and RF voltages on the top gate. The undoped silicon substrate is activated by flashing a surface mounted blue LED to generate free carriers and can then be used as a back gate \cite{Roche2012}.

\subsection{Charge stability diagram}

 We first characterize the device charge stability diagram as a function of top gate ($V_{tg}$) and back gate voltages ($V_{bg}$). Rf-reflectometry is used to detect a change in the sample impedance due to a dissipative or dispersive event occurring below the top gate.  As a consequence it can probe a change of resistance, due for instance to a charge tunneling between the source and a localized state~\cite{Gonzalez-Zalba2015, Colless2013}, or a change of quantum capacitance induced by a charge tunneling between two quantum dots \cite{Petersson2010, Chorley2012}. Fig.~\ref{fig:Sample_stability}(c) presents the filtered numerical derivative of the reflected resonator signal amplitude as a function of $V_{tg}$ and $V_{bg}$. It shows a set of charge transitions with a small dependence on $V_{bg}$ corresponding to single charge tunneling from the source or drain to quantum dots localized below the top gate. We attribute these charge transitions to the so-called corner quantum dots, formed in the top corners of the nanowire where the electric field is maximum, when the channel is in the subthreshold regime~\cite{doi:10.1021/nl500299h}. These dots have relatively large charging energies, $E_c  \sim 18\,$meV, and a strong top gate lever arm, $\alpha_{tg} \sim 0.85$, similar to what has been reported elsewhere \cite{Gonzalez-Zalba2015, doi:10.1021/nl500299h}. However, one single transition, indicated with a black arrow in Fig.~\ref{fig:Sample_stability}(c), is more coupled to the back-gate than the other charge transitions: this corresponds to a localized state lying further away from the nanowire-top gate interface. 

Such subthreshold resonances are commonly attributed to single donors located in the channel~\cite{Verduijn2014, Dupont-Ferrier2013}, which is likely to be the case in the present sample, as the charging event occurs very close to the threshold indicating that the impurity state is close to the conduction band edge, as expected for phosphorus atoms in silicon \cite{Tan2010}. In addition, from the stability diagram and the back-gate/top-gate lever arm ratio, we can place the impurity far below the surface, eliminating any interface charge trap or electron puddle, while in the same time the number of bulk defects in the nanowire active region ($30\,$nm x $30\,$nm x $11\,$nm) is a few orders of magnitude lower than the average number of dopants that were implanted ($\sim$5 dopants in the channel). Finally, no subthreshold signatures were observed in any of the undoped devices (see Ref.~\cite{Gonzalez-Zalba2015}).
 
\section{Charge dynamics} 
\subsection{Interdot charge transition and quantum capacitance}

\begin{figure}[t]%
\includegraphics[width=\columnwidth]{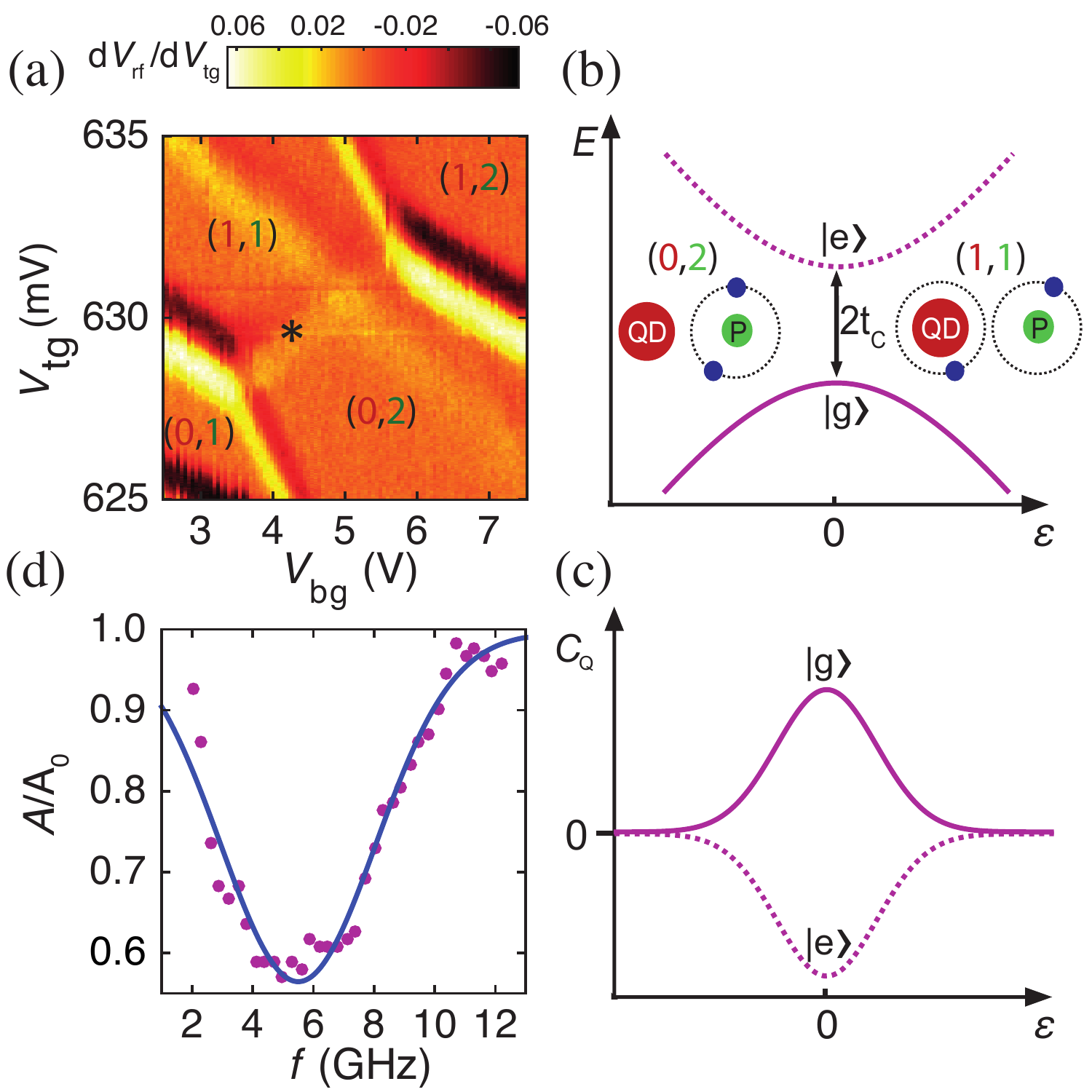}%
\caption{(a) A close-up of the charge stability diagram shows the charge transition between a corner dot and a donor, with charge occupancies labeled as (corner dot, donor). The * indicates the avoided level crossing sketched in (b) the energy levels of the hybrid double quantum dot, where the quantum dot and the P atom form bonding, $\ket{g}$, and antibonding $\ket{e}$ states when the detuning, $\epsilon$, is near zero. For simplicity, spin effects are omitted here and will be discussed later. For $|\epsilon|\gg0$, the double dot has ionic-type wavefunctions where the charges are localized on the dot or donor. (c)~States $\ket{g}$ and $\ket{e}$ have opposite values of quantum capacitance and hence (d) monitoring the interdot charge transition (*), as a function of applied microwave frequency, shows a reduction in signal amplitude upon resonance with the $\ket{g}$:$\ket{e}$ transition (see S3 for more details). 
}
\label{fig:Qcap}%
\end{figure}

In Fig.~\ref{fig:Qcap}(a) we focus on a particular region of interest in the charge stability diagram, showing the classic signature of a double quantum dot through the presence of an extra ridge at the intersection between the charge transitions of a corner dot and the donor. The charge number (1,1) corresponds to one electron located on the corner dot and one at the donor site, and (0,2) indicates that both electrons are on the donor. This charge assignment is deduced from two complementary measurements. First, we exploit the phase contrast of the reflected signal on the stability diagram (see supplemental information, Fig. S1), to assign the two corner dots' charge occupancy. We deduce that the corner dot transition, seen in Fig.~\ref{fig:Qcap}(a), corresponds to a transition from (0,1) to (1,1). Second, from a magnetic field dependence of the donor charge transition, we attribute the charge degeneracy to a $D^0$:$D^-$ transition (see supplemental information, Fig. S2), where $D^0$ is the neutral donor and $D^-$ the anion. Moreover, we observe this transition very close to the threshold, as expected~\cite{VoisinPRB}. 

We now investigate the coupling between the donor and one of the corner dots. The reflectometry signature of a charge tunnelling between the donor and a corner dot is obtained by measuring the resonator response at an intersection between their charge transitions. The  signal amplitude \textit{A} at the ICT can be shown to be a function of the quantum capacitance, its model is described in ref.\cite{Schroer2012} and depicted in Fig.~\ref{fig:Qcap}(b,c). When the energy levels of the donor and the corner dot are brought into resonance, the tunnel coupling between the two quantum systems gives rise to a set of molecular orbitals: a bonding and an anti-bonding state.  The quantum capacitance is directly proportional to the curvature of the eigenenergies with respect to detuning~\cite{Petersson2010},
\begin{equation}
C_q=-(e\alpha)^2\frac{\partial^2E}{\partial\epsilon^2},
\label{Qc}
\end{equation}
where  $\alpha$ is the coupling between the resonator and the double quantum dot, $E$ the eigenenergies and $\epsilon$ the detuning energy. The quantum capacitance is then maximum for $\epsilon = 0$, where the curvature of the eigenenergies is maximum. Notably, the quantum capacitances of the bonding state, $\ket{g}$, and antibonding state, $\ket{e}$, are of opposite sign. 

\subsection{Tunnel coupling and charge dephasing time}
We exploit these quantum capacitance signatures to characterize the tunnel coupling between the donor and the corner dot as well as the charge dynamics in the system, starting by performing MW spectroscopy using a local antenna coupled to the top gate. Fig.~\ref{fig:Qcap}(d) shows the relative change of signal, $A/A_0$, at the ICT as a function of the MW frequency, where $A_0$ is the signal amplitude without any MW applied. The MW excitation causes a fraction of the ground state population to be promoted to $\ket{e}$, reducing the averaged quantum capacitance. As a result, the maximum change of signal occurs when the MW frequency matches the tunnel splitting, which in our case gives $\Delta=2t_c=5.5\,$GHz. We use a Gaussian function to fit the data, from which we extract the charge dephasing time $T_2^*\sim 200\,$ps, very similar to the one measured in a double-donor system in the same type of nanowire transitors \cite{Dupont-Ferrier2013}. The present dephasing rate is close to the tunnel splitting, therefore implementation of quantum protocol that uses coherent charge tranfer to mediate spin information, such as coherent transfer by adiabatic passage (CTAP)\cite{Hollenberg2006}, would have limited fidelity. However, while the tunnel splitting might be tunable in alternative device architectures~\cite{Dupont-Ferrier2013}, the dephasing time may be improved either by going to lower temperature if it is phonon-limited or by improving the charge stability of the device if it is charge noise-limited.

\subsection{Charge relaxation}

\begin{figure}[t]%
\includegraphics[width=\columnwidth]{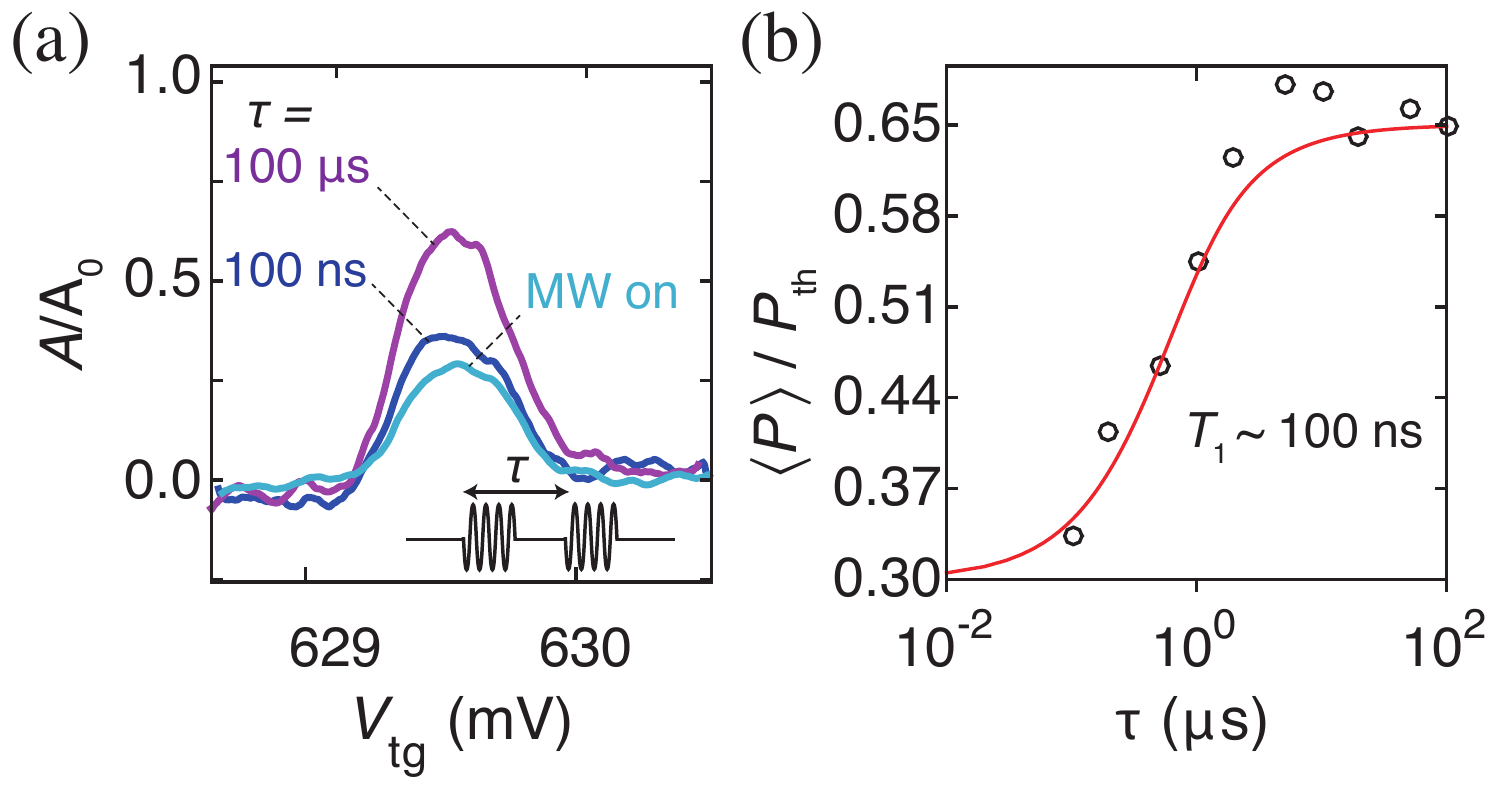}%
\caption{(a) Interdot charge transitions measured under continuous resonant MW (light blue), chopped MW at 10MHz (dark blue) and at 10 kHz (purple). (b) Renormalized average charge state population difference as a function of the chopping period $\tau$. The data are fitted using Eq.~\ref{proba} to give the charge relaxation time \tone.
}
\label{fig:charge_relax}%
\end{figure}
To confirm that the dephasing rate is not limited by phonon-induced charge relaxation, we present measurement of the charge relaxation time, $T_1$, between the bonding and anti-bonding states. We use MW excitation to populate $\ket{e}$ and measure how fast it decays to the ground state. The relaxation time is expected to be  short compared to our demodulator and voltage amplifier bandwidth ($\sim$ a few MHz), making a transient measurement impossible in our case. Instead, we use a procedure developed by Petta \textit{et al.}~\cite{Petta2004}, where the MW excitation is chopped at some frequency $1/\tau$ with a $50\,\%$ duty cycle, while time averaging the signal at the ICT. We define the charge polarisation as $P=(P_g-P_e)$ with $P_g$ and $P_e$ the ground and excited state populations.

When the period $\tau$ is long compared to $T_1$, the time averaged polarisation is $\avg{P}\approx1/2(P_{sat}+P_{th})$. This is because the polarisation is, to a good approximation, at saturation, $P_{sat}$, during the first part of the cycle (MW on) and at thermal equilibrium, $P_{th}$, during the second part (MW off). When $\tau$ is short compared to $T_1$, the system has no time to relax to the ground state and it is then at saturation throughout the cycle, giving $\avg{P} \approx P_{sat}$. Charge relaxation takes the system between these extremes, giving:
\begin{equation}
\avg{P}= \frac{( P_{sat}+ P_{th})}{2} +( P_{sat}- P_{th}) \frac{T_1(1-e^{-\tau/2T_1})}{\tau}.
\label{proba}
\end{equation}

In the present study, the normalized amplitude $A/A_0$ at the ICT  is directly proportional to $\avg{P}/P_{th}$, thus $P_{sat}/P_{th}$ is obtained from the ICT signal under continuous MW excitation (see Fig.~\ref{fig:charge_relax}(a)). The power dependence of the saturation level is plotted in the supplemental information, Fig. S3. Also shown in Fig.~\ref{fig:charge_relax}(a) is the ICT under MW excitation chopped with $100\,$ns and $100\,\mu$s time periods. As expected, the signal amplitude is greater when the charge has more time to relax.

The time averaged polarisation recorded for different values of $\tau$ is presented in Fig.~\ref{fig:charge_relax}(b) and is fitted to give $T_1\sim100\,$ns. Since $T_1\gg T_2$, charge coherence is not limited by coupling with phonons. Therefore, background charge fluctuations or noise in the gate voltages are expected to be the main dephasing sources as they induce fluctuations in both tunnel coupling and detuning energy~\cite{Itakura2003}.

\section{Spin blockade}
\begin{figure}[t]%
\includegraphics[width=\columnwidth]{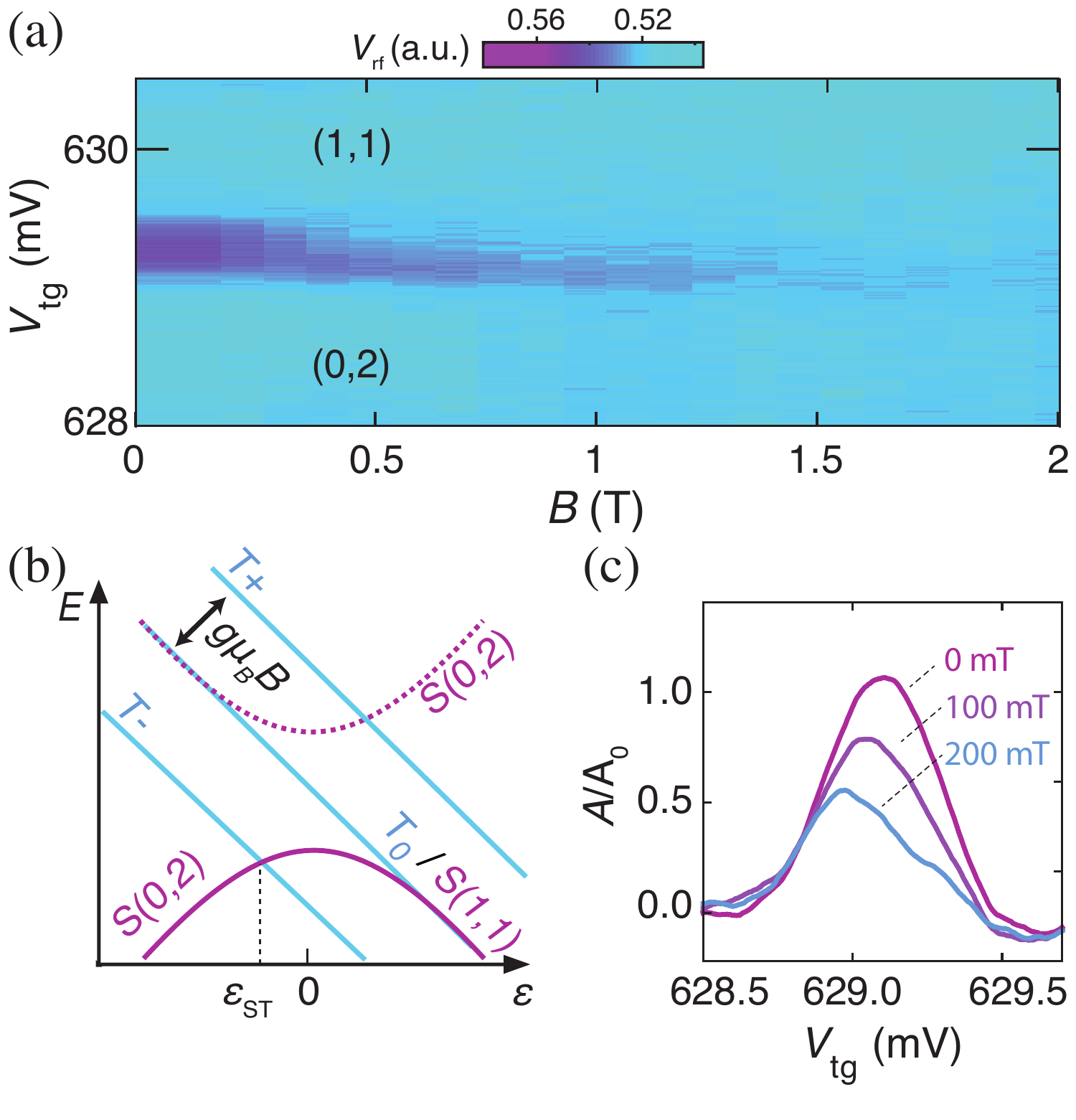}
\caption{(a) The quantum capacitance signature of the interdot charge transition (ICT) become strongly suppressed with increasing magnetic field, as the T$_-$  triplet state becomes the ground state, as shown in (b) the energy level diagram near the (1,1) to (0,2) transition in the presence of a magnetic field. 
$\epsilon_{ST}$ denotes the value at which the singlet and T$_-$ states intersect, and is responsible for the shift and asymmetry observed in (c), the
ICT measured at $0\,$mT, $100\,$mT and $200\,$mT. 
}
\label{fig:SB}%
\end{figure}
We now investigate spin-related effects in the system, in order to demonstrate unambiguously that the ICT is of even parity. Fig.~\ref{fig:SB}(a) displays the ICT as a function of $V_{tg}$ and magnetic field, $B$. It shows that the reflectometry signal disappears with increasing magnetic field and that above $\sim1.2\,$T the variation of quantum capacitance at the ICT has completely vanished. This may be explained by the fact that i) at $\epsilon=0$, as the magnetic field is increased, the population of the singlet state decreases while the population of the T$_-$ triplet state increases (and eventually dominates for $g\mu_BB \gg k_BT$); and ii) the triplet state has a linear dependence on detuning and hence zero quantum capacitance signature according to Eq.~\ref{Qc}. A similar response has been observed in an InAs double quantum dot coupled to a rf resonator \cite{Schroer2012} and has been theoretically investigated in Ref.~\cite{Cottet2011}. 

The shift in the maximum amplitude position observed at low field (see Fig.~\ref{fig:SB}(c)) can then be understood by considering the detuning value at which the singlet and T$_-$ states intersect, $\epsilon_{ST}$, 
which shifts with magnetic field, in accordance with Fig.~\ref{fig:SB}(b). For $\epsilon<\epsilon_{ST}$, there is a singlet ground state and the quantum capacitance has a finite value, albeit one which decreases for more negative $\epsilon_{ST}$. For $\epsilon>\epsilon_{ST}$ there is a T$_-$ ground state and the quantum capacitance vanishes. The asymmetric lineshape observed is a signature that the system has a significant population which follows the ground state, (i.e.\ at the S:T$_-$ intersection) and may be understood by the presence of an avoided level crossing combined with the fact that these measurements are inherently multi-passage experiments. The mixing between these two states is generally induced by the magnetic field gradient created by the nuclear spin bath \cite{Foletti2010}. In contrast,  in this silicon hybrid double dot, the mixing should be dominated by the hyperfine interaction with the nuclear spin localized at the donor site. As a result, neglecting the Overhauser field created by the  $^{29}$Si, the electron located on the donor  is detuned by $\delta = A\textbf{S}.\textbf{I}$ where \textit{A} is the hyperfine coupling constant, \textbf{S} the electron spin and \textbf{I} the donor nuclear spin. In the case of phosphorus, I=1/2 and $A=117\,$MHz, the S(1,1)-T$_0$ coherent evolution at very low exchange coupling could be driven at $\sim 50\,$MHz. In the case of bismuth, I=9/2 and $A=1.48\,$GHz, the driving frequency should be even higher ranging from $300\,$MHz up to $3\,$GHz depending on the nuclear spin state.

%
\section{Outlook and conclusion}
In conclusion, we have investigated a hybrid double dot system formed by a single donor and a corner dot in a single silicon nanowire transistor. Combining rf-reflectometry with microwave spectroscopy has allowed us to determine a tunnel splitting $2t_c=5.5\,$GHz as well as characterize the charge dynamics. The charge dephasing rate $1/T_2^*\sim 5\,$GHz is similar to the tunnel splitting, and could be reduced by removing the nitride spacers in the device, known to possess a large trapped charge density. Also the tunnel coupling might be tuneable in a different architecture, with a split top-gate for instance, such as that used in Ref.~\cite{Dupont-Ferrier2013}.  Furthermore, we have demonstrated spin blockade at the interdot charge transition due to the presence of singlet and triplet states.  The time evolution of such a singlet-triplet qubit should be governed by the donor nuclear spin, enabling controlled rotation gates when combined with NMR excitation~\cite{Kalra2014}. Finally, exploiting the interdot exchange coupling would allow a SWAP operation between the corner dot and the donor atom spin state and eventually its storage in the donor nuclear spin using electro-nuclear double resonance techniques.


 We thank Nicholas Lambert and Andrew Ferguson for discussions. The samples presented in this work were designed and fabricated by the TOLOP project partners, http://www.tolop.eu. This research is supported by the EPSRC through the Materials World Network (EP/I035536/1), ARC Centre of Excellence for Quantum Computation and Communication Technology (CE110001027) and UNDEDD project (EP/K025945/1) as well as by the European Research Council under the European Community’s Seventh Framework Programme (FP7/2007-2013) through grant agreements No. 279781 (ERC) and 318397. C.C.L. is supported by the Royal Commission for the Exhibition of 1851 and J.J.L.M. is supported by the Royal Society.

\bibliography{library}

%


\widetext
\clearpage
\begin{center}
\Large\textbf{Supplementary information for ``Charge dynamics and spin blockade in a hybrid double quantum dot in silicon"}
\end{center}
\setcounter{equation}{0}
\setcounter{section}{0}
\setcounter{figure}{0}
\setcounter{table}{0}
\setcounter{page}{1}
\renewcommand{\thefigure}{S\arabic{figure}}

\section{ Stability diagram charge assignment and magnetic field dependence}
\begin{figure}[ht!]%
	\includegraphics[width=\textwidth]{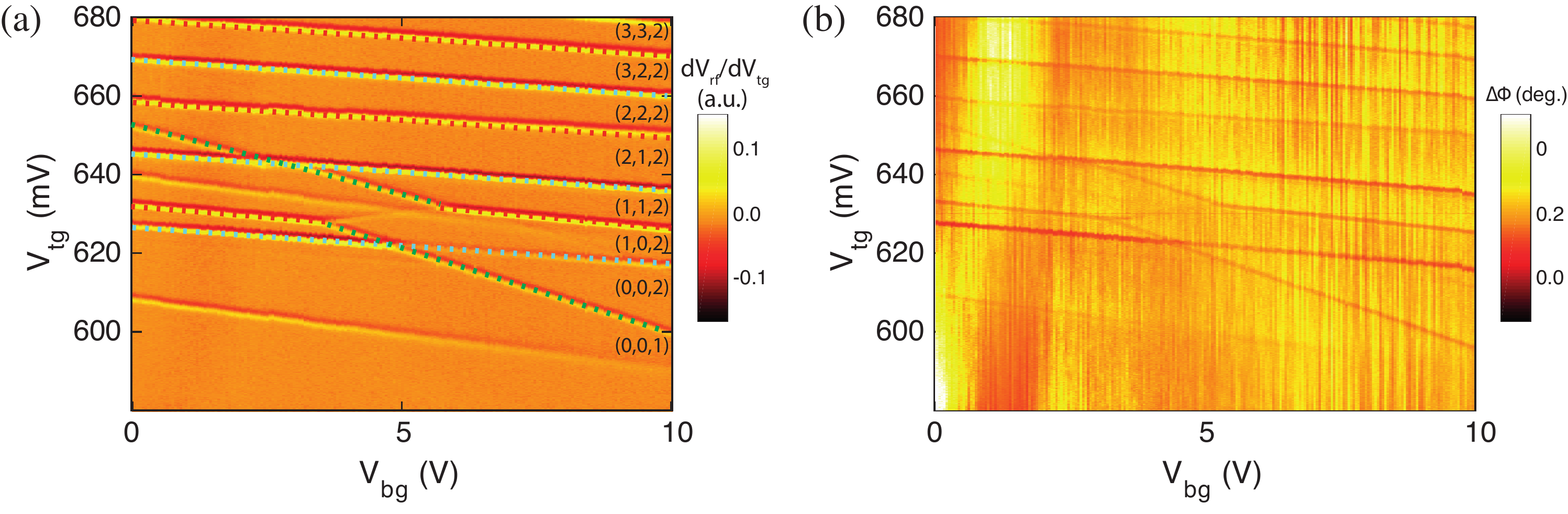}%
	\caption{(a) Numerical derivative, filtered, of the amplitude response as a function of V$_{tg}$ and V$_{bg}$. The last digit of the charge assignment corresponds to the donor atom (green dashed line), the second digit to the coupled corner dot (red dashed line) and the first digit to the uncoupled corner dot (blue dashed line). The unassigned transitions are attributed to another donor in the channel. (b) Phase response of the coupled dot-donor system as a function of V$_{tg}$ and V$_{bg}$. The visibility of the phase response allows discrimination between the two corner dots.}
	\label{fig:phase}%
\end{figure}

RF reflectometry readout is sensitive to the charge dynamics in the double quantum dot and allows us to map out the stability diagram. We can extract the phase response of the device from the I and Q channels. Figure \ref{fig:phase} shows the charge stability diagram of the donor-dot system, with the phase response data plotted on the right. The signal intensity in a single channel does not vary enough to allow us to differentiate between the systems under study. However, the phase response data shows a strong variation in intensity between the different quantum dots in the channel. The degree of visibility of its resonances in the phase response should be characteristic of a particular quantum dot, since it is related to the proportion of its signal in the dissipative versus dispersive regimes. Using this method of discrimination, we are able to isolate three sets of lines in Figure S1(a); a corner dot uncoupled to the donor (blue dotted line), a corner dot that couples to the donor (red dotted line) as well as the donor charge transition (green dotted line).

\begin{figure}[ht!]%
	\includegraphics[scale=0.8]{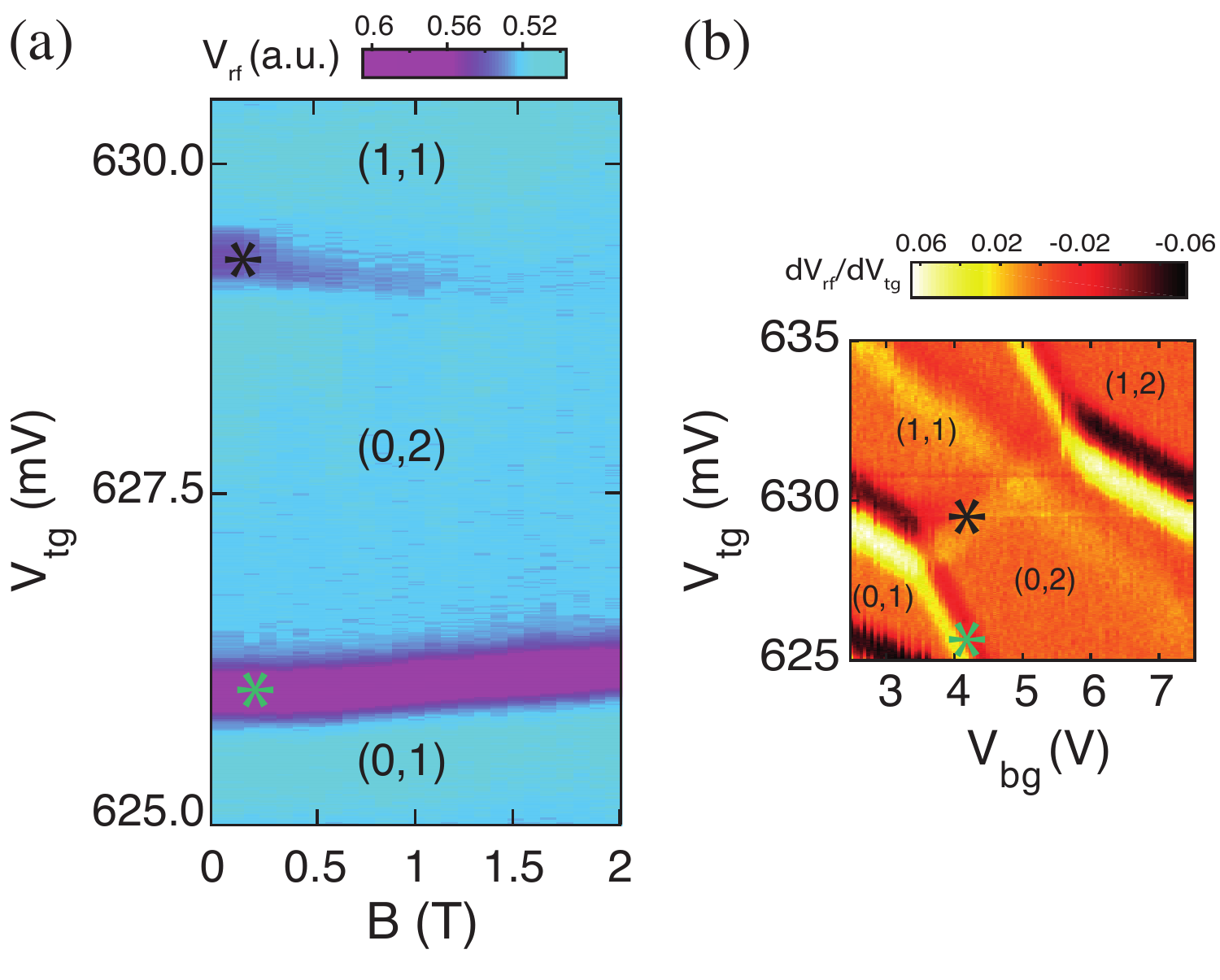}
	\caption{(a) Resonator response as function of of V$_{tg}$ and magnetic field \textit{B}, and (b) as of of V$_{tg}$ and V$_{bg}$. The two charge transitions of interest are labelled with stars in (b). 
	}
	\label{fig:B_dep}%
\end{figure}

Figure \ref{fig:B_dep} shows the magnetic field dependence of the interdot charge transition (ICT) at $\epsilon=0$ as well as of the dopant line. The interdot quantum capacitance signal has a strong field dependence, explored in the main text. As observed in other studies \cite{Schroer2012}, the quantum capacitance signal under Pauli Blockade rapidly diminishes in intensity at this even parity transition under increasing magnetic field, and it also shifts to more negative values of detuning. This allows us to assign the (1,1) and (0,2) states based on the shift in detuning since the T$_-$-singlet crossing shifts in the direction of the (0,2) state. Additionally, while the charge assignment in the stability diagram commences from zero electrons for clarity (at a point in top gate voltage below which we do not see any more dot transitions), in essence we can assert only that we are at an even parity transition at the ICT. 

The dopant line does not diminish under field, but it exhibits a gradual shift towards higher top gate voltages. In general from this we can infer the spin polarity of the tunneling electrons: a shift towards higher top gate voltage, or higher energy, indicates tunneling of spin-up electrons. In our system, as studied previously \cite{Sellier2006, Tan2010}, this shift towards higher energy is consistent with a D$^-$ state, in which electrons form a two-electron singlet. Finally, the D$^+$:D$^0$ charge transition of the donor is not visible in our system. This may be explained by a small tunnel rate between the donor and the reservoir. Indeed, the D$^+$:D$^0$ is deeper below the conduction band than the D$^0$:D$^-$, and as a result, the tunnel barrier between the reservoir and the donor is more opaque \cite{Sellier2006}. Due to the fact that our RF reflectometry technique is sensitive to charges tunneling at a frequency comparable to the excitation \cite{Gonzalez-Zalba2015a}, we are unlikely to see the D$^+$:D$^0$ charge transition.

\section{ Microwave spectroscopy power dependence}
\begin{figure}[ht!]%
	\includegraphics[scale=0.8]{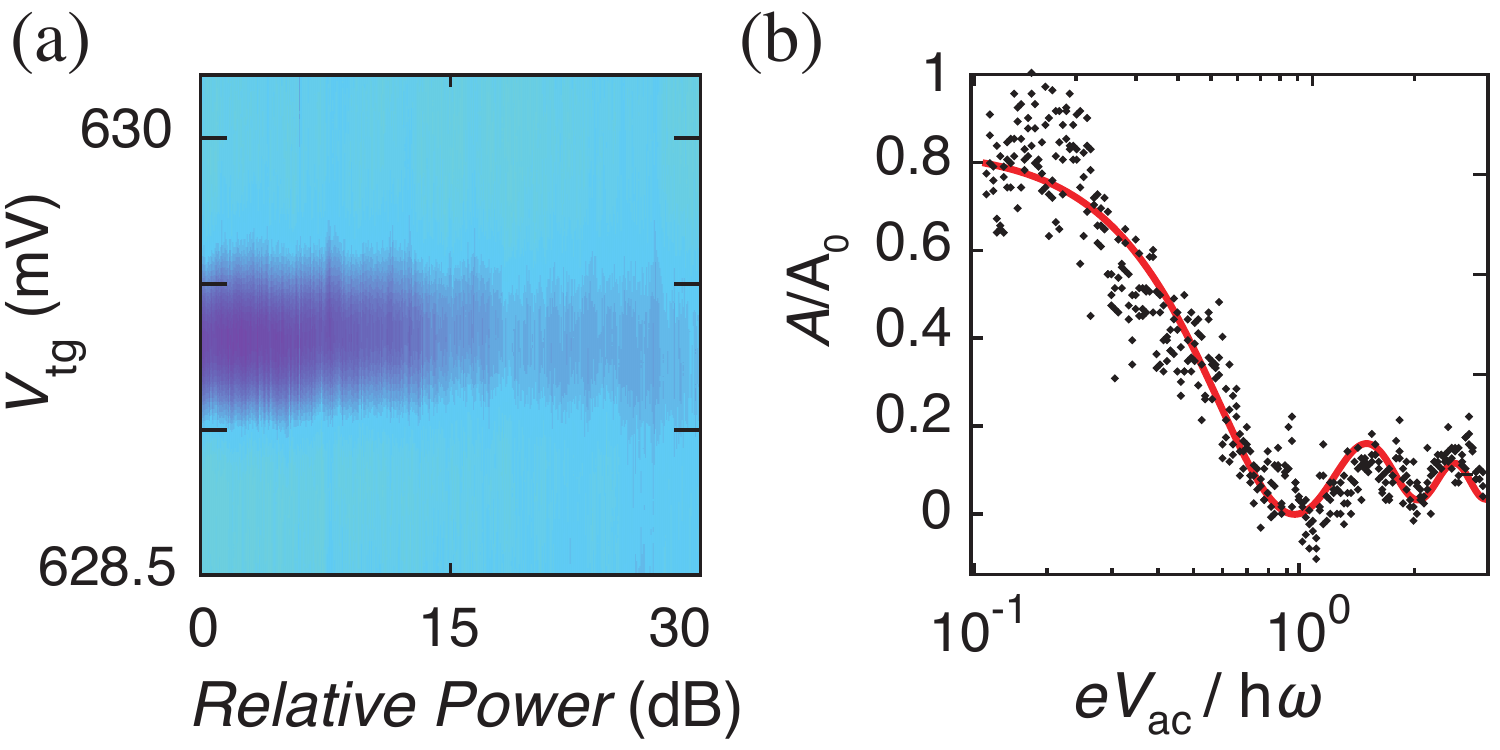}%
	\caption{ (a) Resonator response as a function of V$_{tg}$ and the relative microwave power broadcasted by the antenna. (b) Amplitude of the interdot charge transition as a function of the microwave amplitude. Equation \eqref{PowerDep} is used to fit the data.
	}
	\label{fig:P_dep}%
\end{figure}

Figure \ref{fig:P_dep} shows the interdot charge transition under a continuous microwave excitation of 5GHz as a function of power and detuning. 

In a two level system, Landau-Zener-St{\"u}ckelberg (LZS) interferometry theory provides us with the respective slow passage ($A\nu\lesssim\Delta^2$) and fast passage ($A\nu\gg\Delta^2$) conditions, where $A$ is the driving amplitude, $\nu$ is the frequency, and $\Delta$ is the energy splitting between the excited and ground states. As we vary the power we supply to the microwave antenna, we are varying the driving amplitude $A$, which in turn takes us from the slow to fast passage regime. This, as shown in Figure \ref{fig:P_dep}, gives rise to the expected squared Bessel function behaviour at zero detuning for multiple passage, as theoretically investigated in Ref.~\cite{Shevchenko2010}.
The MW driving tunes the charge polarisation as:
\begin{equation}
P_g-P_e=J_0^2(\alpha)
\label{PowerDep}
\end{equation}
where $J_0^2(\alpha)$ is the square of the 0-th order Bessel function evaluated at $\alpha=eV_{ac}/h\nu$. We send microwaves through an antenna close to the sample, which is then assumed to vary our gate voltage by $V_{ac}$. Our data, while fitting well to the Bessel function, also allows us to calibrate the power arriving at the sample by measuring the distance between the zeros of $J_0^2$, occurring at $\alpha_1=2.4$ and $\alpha_2=5.5$. Figure S3(b) shows the calibrated result.

While we do see a characteristic power dependence at detuning $\epsilon=0$, no higher order photon assisted tunneling has been observed in the high frequency limit ($hf > 2t_c$). This can be explained by considering spin effects in the double dot at the (0,2) to (1,1) transition: for negative detuning, the ground state is the singlet S(0,2) while the excited state is a mixture of triplet states (T$_0$, T$_-$ and T$_+$) and singlet S(1,1). As a result, a MW induced transition between these two states requires a spin-flip process ($\Delta m=1$). Hence, in the absence of spin orbit coupling \cite{Schreiber2011}, or a particular electron-phonon interaction \cite{Colless2014}, MW photon absorption is forbidden.

\section{ Supplementary device}
\begin{figure}[ht!]%
	\includegraphics[scale=0.8]{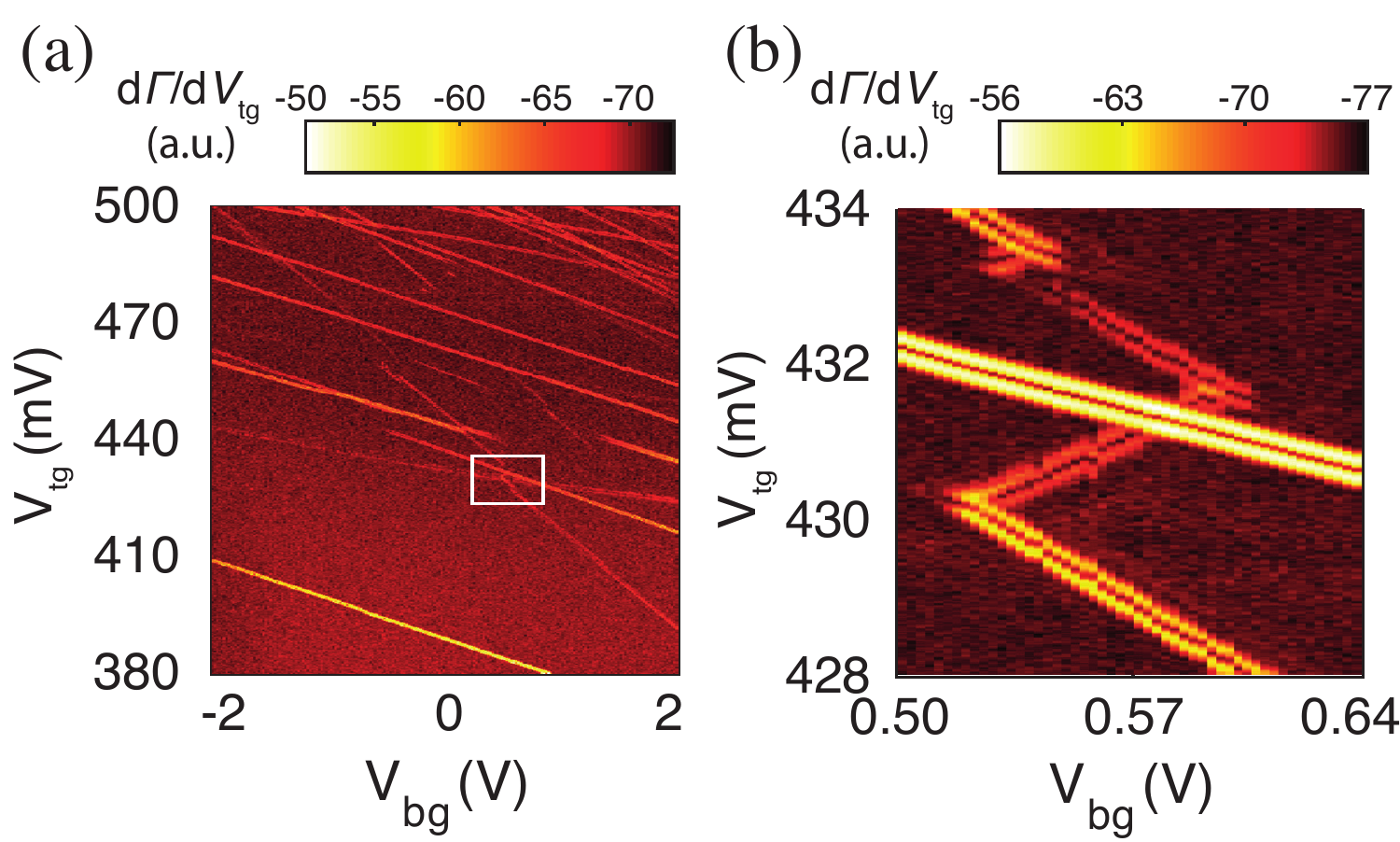}%
	\caption{ (a) Modulus of the derivated reflected amplitude as a function of V$_{tg}$ and V$_{bg}$. The white rectangle corresponds to (b) a close-up of the stability diagram showing a strong signal based on a interdot charge transition. The extra line crossing the interdot charge transition is attributed to an uncoupled corner dot.
	}
	\label{fig:unsw}%
\end{figure}

Figures \ref{fig:unsw}(a) and (b) present measurements done on a similar device as the one presented in the main text, but with a wider and larger channel ($240\,$nm x $100\,$nm x $10\,$nm). A tank circuit with a resonant frequency of $671\,$MHz is connected to the top gate. A small AC excitation is applied to the drain at frequency of $1.1\,$kHz. The amplitude of the resonant side band ($671.0011\,$MHz) is measured as a function of top gate and back gate using a spectrum analyzer. The presence of numerous charge transitions in Fig.\ref{fig:unsw}(a) is attributed to the larger number of dopant in the channel compared to the  device presented in the main text. Fig. \ref{fig:unsw}(b) shows a close-up of the stability diagram  where a donor-based interdot charge transition is visible, similar to what is observed in Fig 2(a).

\end{document}